# A Machine Learning Approach To Anomaly-Based Detection On Android Platforms

Joshua Abah[1], Waziri O.V[2], Abdullahi M.B[3], Arthur U.M[4] and Adewale O.S[5]

[1,3,4,5]Departement of Computer Science,
Federal University of Technology Minna, Nigeria.
[2]Departement of Cyber Security Science,
Federal University of Technology Minna, Nigeria.

## ABSTRACT

*The emergence of mobile platforms with increased storage and computing capabilities and the pervasive use of these platforms for sensitive applications such as online banking, e-commerce and the storage of sensitive information on these mobile devices have led to increasing danger associated with malware targeted at these devices. Detecting such malware presents inimitable challenges as signature-based detection techniques available today are becoming inefficient in detecting new and unknown malware. In this research, a machine learning approach for the detection of malware on Android platforms is presented. The detection system monitors and extracts features from the applications while in execution and uses them to perform in-device detection using a trained K-Nearest Neighbour classifier. Results shows high performance in the detection rate of the classifier with accuracy of 93.75%, low error rate of 6.25% and low false positive rate with ability of detecting real Android malware.*

## 1. INTRODUCTION

Mobile devices have drastically become a ubiquitous computing and storage platform, among these devices; Android holds a large percentage of the market share [1]. In 2013 over 967 million units of smartphones were sold to consumers worldwide, of these smartphones sold to end users in the final quarter of 2013, almost 78 percent ran on the Android platform [2] this amount to sales of almost 220 million units. Based on unit shipments of these smart devices, Android's market share increased further in 2014 with the company holding over 80 percent of the global smartphone Operating System (OS) market in the first quarter of 2014 according to this report. Similarly, in 2014, sales of smartphones to end users totalled 1.2 billion units, with 28.4 percent increase from 2013 as shown in Table 1 and represented two-thirds of global mobile phone sales.

Table 1: Global Smartphone Sales (Thousand Units) to end users by Vendor in 2014 (Source: [3])

| Company | 2014 Units | 2014 Market Share (%) | 2013 Units | 2013 Market Share (%) |
|---|---|---|---|---|
| Samsung | 307,597 | 24.7 | 299,795 | 30.9 |
| Apple | 191,426 | 15.4 | 150,786 | 15.5 |
| Lenovo | 81,416 | 6.5 | 57,424 | 5.9 |
| Huawei | 68,081 | 5.5 | 46,609 | 4.8 |
| LG Electronics | 57,661 | 4.6 | 46,432 | 4.8 |
| Others | 538,710 | 43.3 | 368,675 | 38.0 |
| **Total** | **1,244,890** | **100.0** | **969,721** | **100.0** |

The availability of smartphones at relatively low prices has led to an accelerated migration of feature phone users to smartphones making the smartphone OS market to experience fast growth in most emerging countries, including India, Russia and Mexico [3]. This trend continued to benefit Android, which saw its market share grow by 2.2 percentage points in 2014, and 32 percent year on year as shown in Table 2 and charted in Figure 1.

Table 2: Global Smartphone Sales (Thousand Units) to End Users by OS in 2014 (Source: [3])

| Platform | 2014 Units | 2014 Market Share (%) | 2013 Units | 2013 Market Share (%) |
|---|---|---|---|---|
| Android | 1,004,675 | 80.7 | 761,288 | 78.5 |
| iOS | 191,426 | 15.4 | 150,786 | 15.5 |
| Windows | 35,133 | 2.8 | 30,714 | 3.2 |
| BlackBerry | 7,911 | 0.6 | 18,606 | 1.9 |
| Other OS | 5,745 | 0.5 | 8,327 | 0.9 |
| **Total** | **1,244,890** | **100.0** | **969,721** | **100.0** |

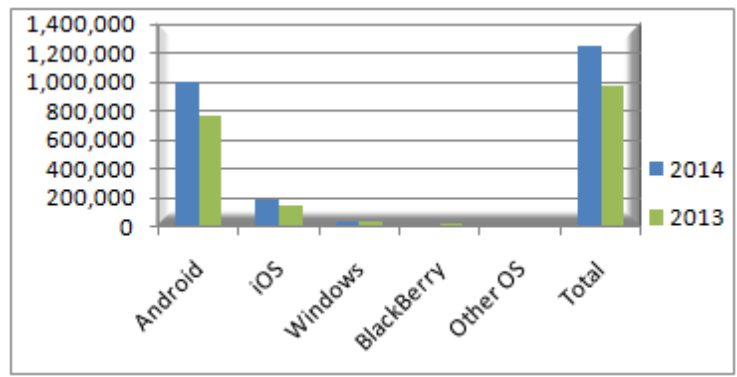

Figure 1: Global Smartphone Sales to End Users by OS in 2014 and 2013 Compared

The ubiquity of the Android platform and in deed the Smartphones in general has not gone unnoticed by malware developers. Mobile devices have become a target to attacks with Android platform being the worst hit [1]. There are already well-known and documented cases of Android malware in both official and unofficial markets [4]. With known malware nefarious capabilities and effects, the detection of malware is an area of major concern not only to the research community but also to the general public. Malware attack is a challenging issue among the Android user community. It therefore becomes necessary to make the platform safe for users by providing defense mechanism especially against malware [5]. Techniques that researchers develop for malware detections are realized through the implementation of malware detectors [6]. Malware detectors are the primary tools in defense against malware and the quality of such detectors are determined by the techniques they employed. Intrusion detection methods can be classified as host-based, cloud-based or social collaboration. In host-based method which is the method adopted in this work, the detection engine runs entirely in mobile device [7]. This rational behind the adoption of this method is the assumption that the capabilities of mobile devices in general and Smartphones in particular increases steadily following Moore's theory [8]. This means that client-server or cloud-based design decisions that moved most of the data analysis processing to the server is changed. The advantages of this architecture are that relocating some of server functionality to the client-side will result in the reduction of communication latencies [9] while the cost in terms of the use of bandwidth is eliminated with ability for real-time detection. The disadvantage of this method is the difficulty in implementation because of the resource-poor limitations of mobile devices but the interest here is purely Smartphones.

Irrespective of the detection method employed, techniques used for detecting mobile malware can be categorized broadly into three categories: Signature-based detection, anomaly-based also known as Behaviour-based detection and Virtual Machine based (VM-based) Detection. Figure 2 shows the relationship between the various types of malware detection techniques.

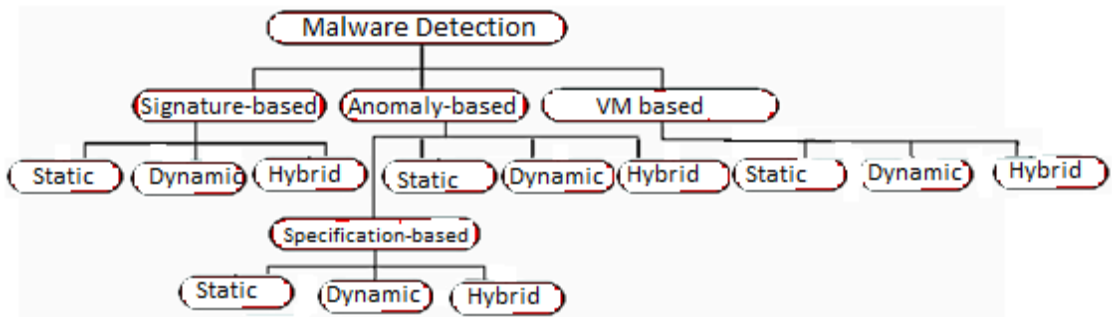

Figure 2: A Classification of Mobile Malware Detection Techniques (Source: [6])

Each of the detection techniques can employ one of the three different approaches: static, dynamic or hybrid (see Figure 2). The specific approach of an anomaly-based or signature-based technique is determined by how information is gathered for use in malware detection. Anomaly-based detection systems use a prior training phase to establish a normality model for the system activity. In this method of detection, the detection system is first trained on the normal behaviour of the application or target system to be monitored. Using the normality model of behaviour, it becomes possible to detect anomalous activities by looking for abnormal behaviour or activities that deviate from the normal behaviour earlier defined occurring in the system. Though this technique look more complex, it has the advantage of being able to detect new and unknown malware attacks. Anomaly-based detection requires the use of feature vectors to train the classifier before subsequent classification can be carried out. These feature vectors are obtained from features or data collected from the system.

This research employs a machine learning approach specifically supervised learning to anomaly-based detection in a host-based manner. A machine learning model is train using a labelled data obtained from the understanding of the application behaviours. The trained classifier is then used to predict future outcomes of test feature vectors. The use of machine learning-based classifier to detect malware poses two main challenges first, the need to extract some sort of feature representation of the applications and secondly, the need for a data set that is almost exclusively benign or well labelled which will be used to train a classifier. In addressing these problems, first a heterogeneous data are extracted processed and vectored into a feature set. Secondly, a K-NN classifier is trained using a normality model that describes to the classifier the normal behaviour. With this, it becomes possible to detect anomalous behaviour by looking for behaviours that deviate from the defined normal behaviours of applications.

This paper is organized as follows: section one gives the introduction; in section two, related works are discussed; in section three, the approach used to realize the design is presented; in section four, the various testing carried out to validate and ensure the workability of the system is explained and the evaluation measures adopted are also defined; in section five, results are presented and discussed accordingly based on the defined evaluation measures defined in section four and finally section six provides conclusions and suggestion for further studies.

## 2. RELATED WORKS

There have been significant research efforts on the problem of mobile malware detection. Generally, malware detection systems employ different approaches. Static analysis approaches such as [9]; [10]; [11] are based on comparing applications to already known malware through a reverse engineering method that decompiles packaged applications and looking for signatures or using other heuristics within the program code. Other approaches like [12]; [13]; [14] monitors the power usage of applications, and report anomalous consumption. [15]; [16]; [17] used a dynamic analysis by monitoring system calls and attempt to detect unusual system call patterns. Some others like [4]; [18] used the universal signature-based approaches that compare applications with known malware or other heuristics. While [19]; [20]; [21]; [22] focused on the use of machine learning and data mining approaches for malware detection.

Although Crowdroid by [16] used a machine learning-based framework that recognizes Trojan-like malware on Android Smartphones, it monitored the number of times a particular system call was issued by an application during the execution of an action that requires user interaction. Crowdroid used about 100 system calls with only two trojanized applications tested. The HOSBAD approach differs from Crowdroid in that it use features extracted from the application layer rather than the kernel layer. Device does not need to be rooted as in the case of intercepting system calls. Furthermore, feature extraction and detection need not be carried out on external servers but in-device in a host-based manner.

Similarly Andromaly by [23] is an intrusion detection system that relies on machine learning techniques. It monitors both the Smartphone and user's behaviours by observing several parameters, spanning from sensor activities to CPU usage. Andromaly used 88 features to describe system behaviours besides rooting the device and the use of external Linux server; the features are then pre-processed by feature selection algorithms. Compared to Andromaly, HOSBAD approach use fewer sets consisting of five features without the need for external server; HOSBAD is host-based and does not require the device to be rooted.

In the work of [24] a machine learning approach was adopted using the clustering model for the analysis of static function calls from binaries to detect anomalies. This technique was used to detect malware on the Symbian operating system. The framework included a monitoring client, a Remote Anomaly Detection System and a visualization component. Remote Anomaly Detection

System is a web service that receives, from the monitoring client, the monitored features and exploits this information, stored in a database, to implement a machine learning algorithm. In HOSBAD, the detection is done dynamically rather than the statically furthermore, the detection will be performed locally and need no connection to remote server saving bandwidth cost, eliminating latency delay and, more importantly, in real-time. HOSBAD uses a supervised machine learning model the K-NN classifier as opposed to clustering being used by Aubrey *et al*. Instead of performing static analysis on the executable to extract functions calls usage using *readelf* command, HOSBAD will use dynamic analysis of applications at the application layer. In the work of [18]; [25] their detection framework were targeted at the Symbian platform while HOSBAD operates on the Android platforms the most used mobile platform in recent times. They discriminate the malicious behaviour of malware from the normal behaviour of applications by training a classifier based on Support Vector Machines (SVM). HOSBAD approach does not only differ from their work in terms of platform but also uses KNN classifier as opposed to SVM used by these authors. The reason is that mobile devices are resource constrained, SVM requires large memory as opposed to KNN classifiers, and this makes KNN more suitable for host-based implementation than SVM.

Dini *et al*., [26] presented Multi-Level Anomaly Detector for Android Malware MADAM. MADAM is a multi-level system that extracts features from both the application level and the Kernel level using a total of 13 features to describe the system behaviour including system calls. It is target on a rooted device and adopted a static approach that requires applications to be decompiled. Unlike HOSBAD approach, MADAM did not consider phone calls made or received by applications in describing system behaviour.

Portokalidis *et al.,* [27] used a VMM approach to malware detection in their design of Paranoid Android system where researchers can perform a complete malware analysis in the cloud using mobile phone replicas. In their work, the phone replicas are executed in a secure virtual environment, limiting their system to no more than 105 replicas running concurrently. In their approach, different malware detection techniques can be applied but compare to HOSBAD approach, there is no restrictions to the number of users since the approach is host-based where the detection will be local to the device without the need for cloud infrastructure benefiting from the advantages of local detection systems.

Mirela *et al*., [28] used neural network approach this proved effective in detecting fraud calls and imposters. The disadvantage of this method is that the process is relatively slow and this method classifies applications into groups having same behaviours and hence, there will be lot of false positives. Jerry *et al.,* [29], focused on viruses that are transmitted through SMS messages and other communication interfaces like Bluetooth and Infrared. But they did not concentrate on worms that will automatically make high rate calls from the mobile device which will incur loss to the user as they are only collecting and monitoring SMS traces.

## 3. APPROACH

HOSBAD is a Host-based Anomaly Detection System targeted at Android Malware propagated via SMSs and calls. HOSBAD integrates data mining with supervised machine learning techniques in its implementation. It is designed to monitor and extract Android's applications data at the application layer and using these data to detect malware infections using a supervised machine learning approach to differentiate between normal and malicious behaviours of applications. The problem of anomaly detection can be considered as a problem of binary classification according to [26], in which good behaviour is classified as "Normal", while abnormal ones are classified as "Malicious". The supervised machine learning model could be viewed as a black box having as input sets of behaviours formatted into sets of feature vectors in ARFF and the output is a flag of "Normal" or "Malicious". The supervised machine learning model which in this case is the K-NN classifier is trained using a normality model on how to classify correctly each element of the feature vector. The training of the classifier takes place at the point called the training phase. This phase is critical because the accuracy of the classifier is dependent on the training phase hence a good training set must be supplied to the classifier. Table 3 shows the list of monitored features used in this work.

To generate a good feature vector that represents typical Android device behaviour HOSBAD utilize features that represents behaviours when the device is active and when it is inactive. However, our training set also contains some malicious behaviour, which strongly differs from the normal ones. Choosing the right features to best represent the device behaviours is a critical task, since their number and correlation determine the quality of the training set [30].

Table 3: List of Monitored Features

| S/No. | Features |
|-------|----------|
| 1. | In/Out SMSs |
| 2. | In/Out Calls |
| 3. | Device Status |
| 4. | Running Applications/Processes |

## 3.1 Malware Detection Processes

HOSBAD combines features extracted from different categories of the device functionality as given in Table 3. First, it monitors the device activities and extracts the features associated with these activities. Secondly, it observes correlation among the features derived from the events belonging to the different activities. In order to extract features first, HOSBAD monitors the in-coming and out-going SMSs. An application may send SMS during its execution and for this to happen, it must contain the SEND_SMS permission in its manifest file otherwise the application will crash as soon as it tries to send SMS message by invoking the SEND_SMS function. The Permissions required by an application are displayed to the user during installation time and the user must either accept all or forfeit the installation of the application that is, the user must agree with all the stated permissions for the application to get installed. This mechanism provides a rough control that can be effective especially to new Android users. An application that gets the SEND_SMS permission could be harmful, since it is able to send SMSs including premium SMSs without the knowledge of the user. This attack is performed by a good number of Android malware that can be found in the markets, since this attack is easy to realize and inimical to the user's credit for this reason, HOSBAD monitors SMS usage.

Furthermore, the system monitors In-coming and Out-going calls since a good number of Android malware initiates calls unauthorized by the user. Lastly, the system monitors the status of the device. The device status is a very good feature since the activity of the device is usually more intense when the user is actively interacting with the device. It is a clear fact that after a very short period of user inactivity the device screen is turned off by the OS, the device can be considered active either if the screen is in 'ON' or there is an active voice call [26]. The elements of the datasets are vectors with $M + 5$ features, where $M$ represents features added to the 5 features during the process of parsing or pre-processing by the parser function, these features are two in number namely; the Date/time stamp and the Applications/processes running on the device at the time. The 5 features include the number of OutCall, InCall, OutSMS, InSMS and the device status (active or inactive). An architectural description of the HOSBAD system is given in Figure 3.

To monitor and extract the stated features from the device, the design includes three monitors; the call monitor, the SMS monitor and the device status monitor see Figure 3. A collector receives these features from all the monitors and then builds the vectors in .csv format. These vectors are parsed and converted to arff; the format acceptable by Weka machine learning tool and stored in local files on the SD card in .arff using a logger module so that they can be used as test set.

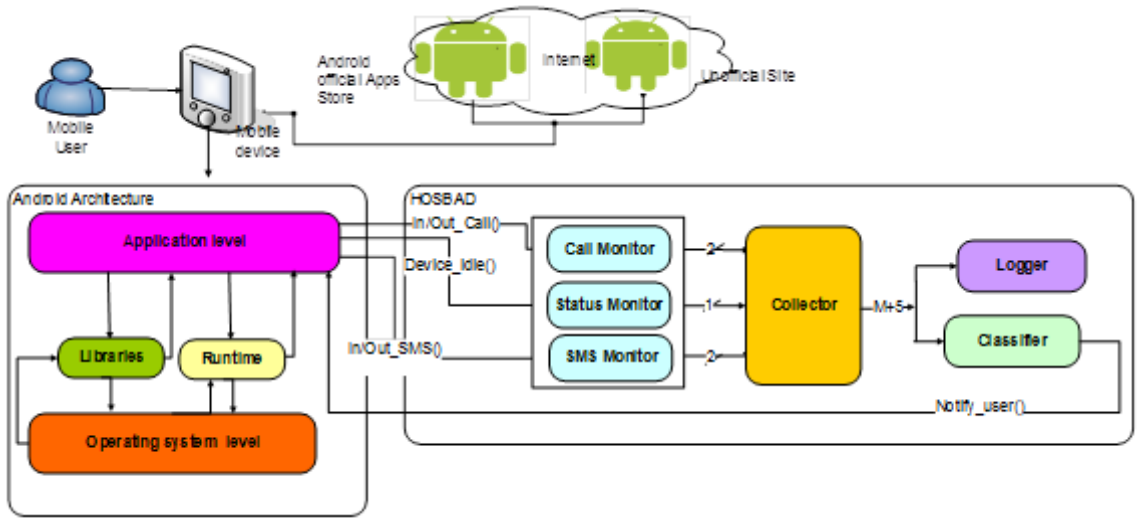

Figure 3: Architecture of the Host-based Anomaly Malware Detection System (HOSBAD)

The classifier module is responsible for performing behaviour-based analysis in which Android applications are classified as either Normal or Malicious. This is done by employing the trained K-NN classifier. A key process of the system is the training phase which identifies the behaviour of the applications. It identifies Android applications into two classifications namely: Normal and Malicious. Figure 4 gives a complete representation of the processes involved in the different phases of the malware detection.

The Waikato Environment for Knowledge Analysis (WEKA) tool [31] is used programmatically by adding the weka.jar file as external jar to the project. After this is done, the machine learning models embedded in the Weka tool become available and are accessed via some sets of Application Programming Interfaces (APIs) provided by Java. The K-NN classifier of Weka tool is invoked and train using set of feature vectors called the normality model.

## 3.2 The Normality Model

In order to efficiently develop a machine learning model, it is important to train the model on the normal and abnormal behaviour of the system. To do this, a normality model is required to describe to the classifier the pattern of behaviours. Hence a normality model is designed based on the fact that malware requires user interaction to activate its payload on the target device. For malware that uses SMS and calls as its propagation vector, it becomes evident that user interaction is necessary for such malware to propagate. SMSs and calls require user interaction with the device to compose and send SMSs or to initiate calls. Therefore, a normal SMS and call activity is one that has active user interaction. In this work, five (5) features were used to describe a normality model for the K-NN model, these features include:

i.     The out-going call
ii.    The In-coming call
iii.   The Out-going SMS
iv.    The In-coming SMS and
v.     The device Status.

These features were used as follows;

i.     If the device is active or inactive at the point of any activity;
ii.    If any SMS is being sent or received when the phone is inactive and
iii.   If any call is being made or received when the phone is inactive

Using a binary representation for these features, the number of probable permutations of these 5 features is obtained by the formula in (1)

$$2^n \qquad (1)$$

Where n is the number of features to be represented.

In this case our n = 5; so we get $2^5 = 32$ instances of the features. The value of 1 represents the presence of the feature while the value of 0 represents the absence of that feature. The numeric count of how many occurrence of the feature is immaterial because even a single presence is enough to describe the entire behaviour. For the device status, 1 represents an active user interaction where the device screen is 'ON' while 0 represents no interaction with the device with the screen turned 'OFF' or hibernated. The combination of the features gives thirteen (13) normal instances and nineteen (19) malicious instances based on the condition that certain activity do not occur at the same time and at certain device state. For example the instances given as;

| Outcall | InCall | outSMS | InSMS | Device Status | Class |
|---------|--------|--------|-------|---------------|-----------|
| 0 | 0 | 1 | 0 | 0 | Malicious |
| 0 | 0 | 1 | 0 | 1 | Normal |

The first instance signifies the occurrence of an out-going SMS while the device screen is in an inactive (OFF/hibernated) state. The application behaviour represented by this instance is suspicious; the reason is that sending SMS requires active interaction with the device; to compose the text message and then send it by pressing the send button. This activity would not have been made by a valid user when the device is idle hence; it is classified as a malicious behaviour. In the second instance, the out-going SMS occurred while the device screen state is active signifying that there is active interaction with the device that kept the screen light ON hence, the activity is classified as a normal activity. This normality model is parsed and converted into Attribute Relation File Format (ARFF) with the date/time stamp and application and or services features appended to each instance and used to train the K-NN classifier.

### 3.3 The Training Phase

The processes of data gathering, pre-processing or parsing and classifier training are collectively referred to as the Training Phase while the stages where the user interacts with the device and its installed applications resulting to monitoring and recording of the feature vectors, parsing of feature vectors, classification and report generation are referred to as the Operative Phase. In the training phase, behaviour model for Android applications represented by the normality model is used to train the classification algorithm. The training or learning phase creates a trained machine learning model or classifier with knowledge of how to classify or predict the future test set that will be supplied to it by the feature extraction module at the operative phase. The normality model is parsed and translated into ARFF for Weka to be able to process the collected data. In the training phase, a training set is used to obtain a trained K-NN classifier while in the operative phase, the user uses the device and each monitored feature vector is given as input to the trained K-NN classifier that classifies them as Normal or Malicious with a notification immediately displayed on the device screen for the user who then decides whether to leave the application or service running in the device or to uninstall it depending on the outcome of the classification result.

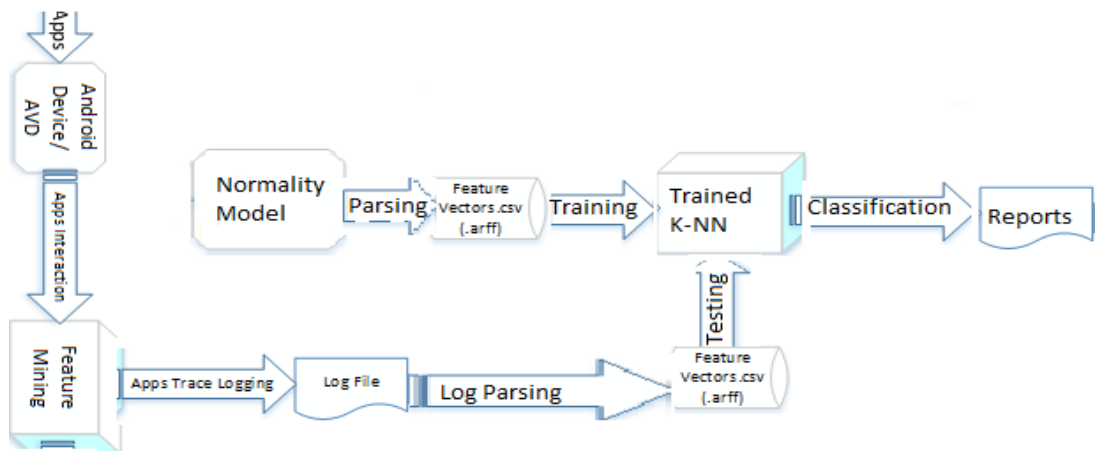

Figure 4: Stages of the Malware Detection System

### 3.4  K-NN Classification

The algorithm used for this module is the K-NN model or classifier which is one of the machine learning models embedded within the Weka tool. This is part of the Operative Phase which combines testing using the test feature set used for the classification after the training phase. The testing feature set is the output of the feature extraction module which is a file in arff containing unclassified instances of application behaviours to be tested or classified. Figure 8 shows a sample of test.arff file generated by the monitoring module to be passed to the classification model for classification. The classifier out this file as a labeled.arff file which is the classified file output from the K-NN classification model.

In the training of the K-NN classification model, samples containing 32 instances (13 normal and 19 malicious) obtained from the normality model were used. In order to get test feature vectors, the device monitoring was carried out by the feature extraction module whose output is arff containing the instances of application behaviours. The number of instances depends on the number of applications installed on the device, the user interactions with the installed applications and the duration of the monitoring. This file is used as test feature which is passed to the trained K-NN classifier for prediction or classification. The evaluations are carried out based on the defined evaluation criteria in (2) to (11) given in section 4.1. A 10-fold cross validation was employed; this strategy was chosen to provide a wider range of samples for the testing of the classifiers' ability to detect unknown malware.

## 3.5 Implementation

The detection framework has been developed on a laptop machine with the Intel Core-i3-370M Processor, 3GB of available memory and 500GB Hard Disk Drive (HDD). This machine runs Windows 7 Operating System and tests are carried out on a TECNO P5, with Android Jelly Bean version 4.2.2 OS, and Linux kernel version 3.4.5. The implementation does not require rooting or jail breaking the device since the monitored features are all carried out at the highest layer; the application layer. The component of the system framework includes an Android Application in Java implemented using the Android Studio version 1.3.2 Integrated Development Environment (IDE) as the Software Development Kit (SDK).

The first component of the Java Application is the Device Monitoring, which monitors the device for activities. These activities include (i) sent SMSs (ii) received SMSs (iii) initiated calls (iv) calls received and (v) the device status (idle/active). The Java application also includes two task modules; application-level logger (see Figure 3), which reads the vectors built by the collector in the .csv format and parse them into .arff and then logs them on the SD card. The second task is the classifier that states if the vectors built by the collector and logged in the logger are normal or malicious. If an instance is classified as malicious, the classifier sends a notification to the device screen for the user and then the classified vector is logged on the SD card. For classification Weka version 3.7.3; an open source library in Java that includes several classification tools was used by adding the Weka.jar file as an external library to the Android Studio project from where the available features are invoked programmatically using sets of available Java APIs. The screen display of the integrated detection system is given in Figure 5.

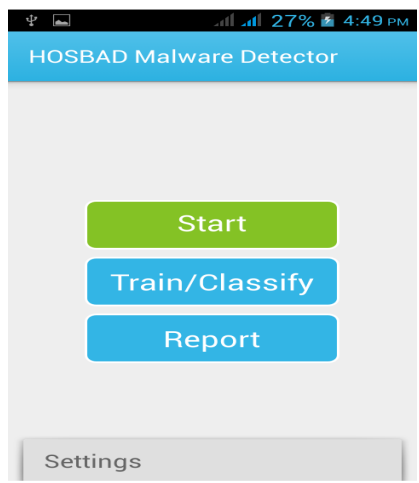

Figure 5: Screen Shoot of the HOSBAD Application

## 4. HOSBAD SYSTEM TESTING

This section discusses the tests that were performed on the system to ascertain its effectiveness and performance. The Application was installed on different devices running different versions of Android and reports are collected and analyzed. In order to have enough dataset for analysis, the design include a means of obtaining monitored reports remotely via email such that the application can be installed on different versions of Android and configured to send reports to the desired email address without the user's interference. Users of these devices will interact with their device normally and after a desired period of time, the monitoring will be stopped and the reports immediately sent to the specified email address. Figure 6 shows the configuration interface of HOSBAD for local and remote reporting. The Results obtained from monitoring the device shown in Figure 8 identifies the tested applications under AppName attribute in the report. These applications are not known prior to monitoring but as the monitoring is done, the applications are added and combined with the values from the Training file to make a new file the test.arff. The normality model which was carefully constructed to represents the complete permutations of the features was used for training the classifier. Tests were carried out on the trained classifier in the operative phase where the user uses the device. During this time, the device monitoring is carried out where features were extracted, parsed and logged into feature vectors in ARFF. The monitoring report was then used to test the performance of the classifier by running the classifier against the reported feature vectors.

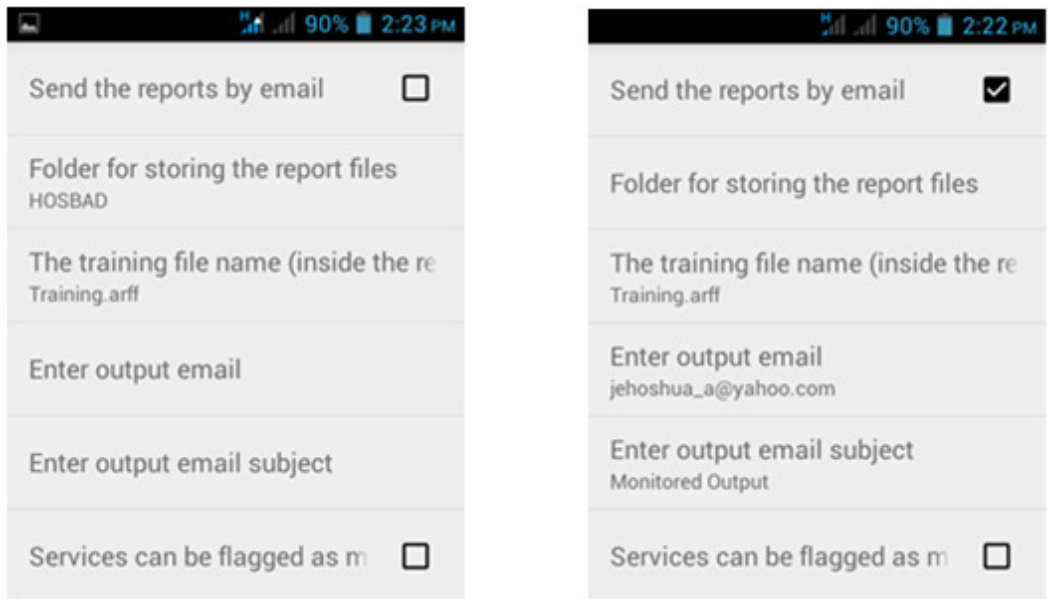

Figure 6: Screen Shoots of the HOSBAD Application Configuration for Local and Remote Reporting System

In a normal usage, SMS and Calls require active interaction and both actions cannot be performed simultaneously that is; SMS cannot be composed and sent at the same time that a call is being initiated. However, some applications send or receive SMSs to provide some kind of services. Since, SMS is a costly service, if compared to the amount of data that are sent with a message, applications should avoid SMS as communication channel as much as possible, and they should require that the user actively agrees with the sending of each message. Applications that send SMS messages or initiates calls when the device is inactive should be considered malicious. Similarly, Applications that sends SMS while an active call is on or vice versa should be classified as malicious. For all these reasons, a log of several SMS sending operations were made and several calls initiated to represent real-life scenarios and the resulting vectors added to the dataset. To properly test this application's performance, an SMS malware called SendSMS that automatically sends SMSs to a specified number at certain time intervals was also developed. This malware sends the SMS without the user being aware and uses up the users' credit without giving message sending report. The feature vector of this SendSMS application is observed in the monitored report feature vector instances and the classifier run on the feature vectors to see if the developed malware will be caught by HOSBAD and classified adequately.

Classifiers are not able to predict suspicious elements if they are not trained. As previously stated, a malicious behaviour is one that strongly deviates from those known to be normal. Hence, we have manually defined some malicious elements by creating vectors with SMS and call occurrences, when the user is idle, and vectors with SMS and calls both initiated at the same time. Figure 8 shows example of a report produced during device monitoring. The first entry in the report is the date/time stamp, the second entry is the running application or service, the third entry is the OutCall, the fourth entry is the InCall, the fifth entry is the OutSMS, the sixth entry is the InSMS. These header titles are defined by the ARFF @ symbols after the @relation key word. The seventh entry is the device status and the last entry which is not derived from the monitoring events represents the class the instance belongs to after classification. Before the classification the value for this field is '?' which is Weka's default way to represent an unknown class, these are replaced immediately after classification is carried out on the file to produce the predicted value of either "Normal" or "Malicious". Figure 8 shows a parsed monitored and extracted feature vector that has not been classified.

To represent malicious behaviours concerning SMS messages, we have manually added to the test set some vectors with a number of sent messages. It is to be noted here that if the classifier is trained using such dataset, which does not include malicious vectors generated by real malware then each malware, if detected, can be considered as a zero-day-attack according to [26]. The classifier is a K-Nearest Neighbours (K-NN) with K = 1 (also known as 1-NN). This classifier has very good performance and can easily adapt to a large number of problems, requiring a small amount of computation time to classify an element and a trivial update algorithm amenable to mobile devices which are limited in their available resources. Figure 7 shows the working display of HOSBAD detection system using features vectors with normal instances and one with malicious instances.

## 4.1 Evaluation Measures

There are many evaluation measures proposed in literature for evaluating the predictive accuracy of machine learning models. Some of these measures have been employed by scholars in their machine learning researches in time past, for example [32]; [21]; [33]. In the context of our problem, the relevant measures used in line with this research are detailed hereafter.

Supposed that $nN \rightarrow Normal$ is the number of Normal applications correctly classified as Normal, and $nN \rightarrow Malicious$ be the number of Normal applications incorrectly classified as malicious, $nM \rightarrow Malicious$ be the number of malicious applications correctly classified as Malicious and $nM \rightarrow Normal$ be the number of malicious applications incorrectly classified as Normal. The accuracy of the classifier's algorithm which is the proportion of the total number of predictions that were correct is given in (2).

$$Accuracy\ (Acc.) = \frac{nN \rightarrow Normal + nM \rightarrow Malicious}{nN \rightarrow Normal + nN \rightarrow Malicious + nM \rightarrow Normal + nM \rightarrow Malicious} \tag{2}$$

The error rate is given by:

$$Error\ rate\ (E.R) = \frac{nN \rightarrow Malicious + nM \rightarrow Normal}{nN \rightarrow Normal + nN \rightarrow Malicious + nM \rightarrow Normal + nM \rightarrow Malicious} \tag{3}$$

The accuracy measure in (2) shows the general statistics of correctly classified instances, whether Normal or Malicious of the machine learning model during the testing phase. The error rate given by (3) can also be calculated by the equation;

$$Error\ rate = 1 - Accuracy \tag{4}$$

This is the complementary measure of the accuracy.

Other evaluation measures like Positive predictive value or precision also known as True Positive Rate (TPR), Negative predictive value or True Negative Rate (TNR), False Positive Rate (FPR), False Negative Rate (FNR) and Precision ($\rho$) are also defined and given in (5) to (11).

$$TPR = \frac{nM \rightarrow Malicious}{nM \rightarrow Malicious + nM \rightarrow Normal} \tag{5}$$

$$TNR = \frac{nN \rightarrow Normal}{nN \rightarrow Normal + nN \rightarrow Malicious} \tag{6}$$

$$FPR = \frac{nN \rightarrow Malicious}{nN \rightarrow Malicious + nN \rightarrow Normal} \tag{7}$$

$$FNR = \frac{nM \rightarrow Normal}{nM \rightarrow Malicious + nM \rightarrow Normal} \tag{8}$$

$$\rho = \frac{nM \rightarrow Malicious}{nN \rightarrow Malicious + nM \rightarrow Malicious} \tag{9}$$

The TPR is also known as the model's detection rate and it is the proportion of positive cases that were correctly identified. It is the number of truly malicious applications that are classified as Malicious divided by the total number of malicious samples in the test set; this measure represents the classifier's ability to detect 'unknown' malicious samples. The FPR, with respect to the malicious class is measured by the number of true Normal applications misclassified as Malicious to the total number of Normal instances recorded during testing. This is complementary to the TNR given by (6). FNR is the measure of the classifier's tendency to misclassify malicious applications as Normal; the FNR measure is complementary to the TPR or detection rate. The precision measure ($\rho$) indicates the precision of the classifier when it takes decision to classify a sample as Malicious. Finally, the Area under the Receiver Operating Characteristics (ROC) curve, (AUC) which is the total area under the plot of FPR against TPR is measured.

Hence,

$$0 \leq AUC \leq 1$$

Given an ROC curve, a perfect classifier will have an AUC of 1. Thus, the closer the AUC is to 1, the greater the classifier's predictive strength and hence the performance. We also have Sensitivity or recall defined as the proportion of actual positive cases which are correctly identified; this is given in (10).

$$\text{Sensitivity (Sen.)} = \frac{\text{nN} \rightarrow Normal}{(\text{nN} \rightarrow Normal + \text{nM} \rightarrow Normal)} \tag{10}$$

Similarly, we define specificity as the proportion of actual negative cases which are correctly identified. This is defined in (11).

$$\text{Specificity (Spec.)} = \frac{\text{nM} \rightarrow Malicious}{(\text{nN} \rightarrow Malicious + \text{nM} \rightarrow Malicious)} \tag{11}$$

## 5. RESULTS AND DISCUSSION OF RESULTS

Figure 7 shows screen shoots of HOSBAD after monitoring and detection was carried out first on feature vector without any malicious instance and then on feature vector with two malicious instances. Detail result from monitoring and detection are given in section 5.1.

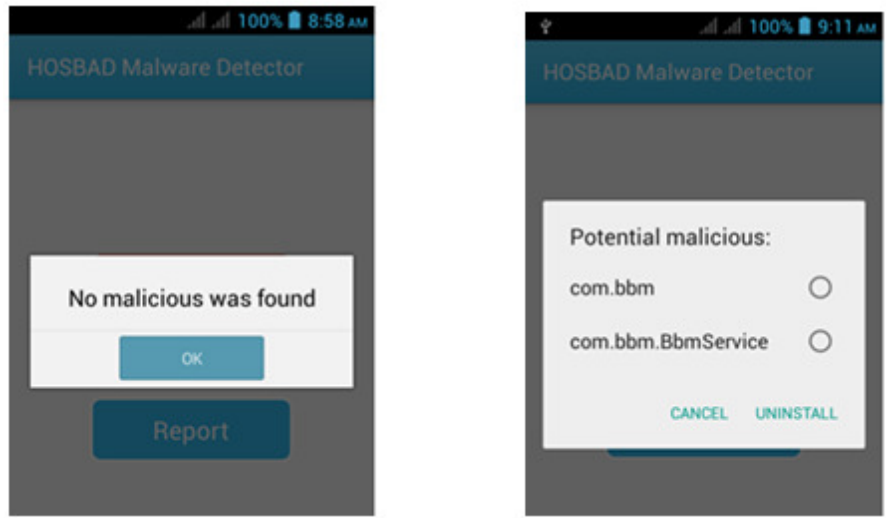

Figure 7: Screen Shoots for Normal and Maliciously Classified Instances

### 5.1 Results

The results obtained from the test are shown in Figure 9 and clearly tabulated and summarized in Table 4. These results were obtained from a single run of the detection model and are discussed based on the evaluation measures earlier discussed in section 4.1.

```
@relation AppFeatureVectors
@attribute Time date 'MM.dd.yyyy HH:mm:ss'
@attribute AppName {'File Manager','Android
System','File','Manager','com.whatsapp.c2dm.C2DMRegistrar','com.facebook.conditionalworker.Con
ditionalWorkerService','com.facebook.vault.service.VaultManagerService','com.facebook.fbservice.s
ervice.DefaultBlueService','com.bbm','com.google.model.mi','com.android.launcher','com.whatsapp'
,'com.android.wp.net.log','ua.com.doublekey.devicemonitoring','com.android.mms','com.android.co
ntacts','com.facebook.platform.common.service.PlatformService','com.mediatek.voicecommand.ser
vice.VoiceCommandManagerService','com.facebook.push.mqtt.MqttPushService','ua.com.doubleke
y.devicemonitoring.ServiceMonitoring','com.tencent.mm.booter.CoreService','com.sys.battery.contr
ol.MainAccessibility','com.glympse.android.hal.GlympseService','com.mediatek.CellConnService.Pho
neStatesMgrService','com.mediatek.filemanager.service.FileManagerService','com.yeahmobi.androi
d.push.impl.PushService','com.whatsapp.messaging.MessageService','com.mediatek.FMRadio.FMRa
dioService','com.facebook.selfupdate.SelfUpdateFetchService','com.whatsapp.VoiceService','com.ye
ahmobi.android.push.impl.CollService','com.bbm.BbmService','com.service.mainService','com.aqpla
y.sdk.AqService','adobe air','MonkeyTest','BBM','Launcher','System UI','Settings'}
@attribute OutCall {0,1}
@attribute Incall {0,1}
@attribute OutSMS {0,1}
@attribute InSMS {0,1}
@attribute Screen {0,1}
@attribute Class {Normal,Malicious}
@data
'10.03.2015 08:40:38','com.bbm',1,0,1,1,?
'10.03.2015 08:40:38','com.google.model.mi',0,0,0,0,1,?
'10.03.2015 08:40:38','com.android.launcher',0,0,0,0,1,?
'10.03.2015 08:40:38','com.whatsapp',0,0,0,0,1,?
'10.03.2015 08:40:38','com.android.wp.net.log',0,0,0,0,1,?
'10.03.2015 08:40:38','ua.com.doublekey.devicemonitoring',0,0,0,0,1,?
'10.03.2015 08:40:38','com.android.mms',0,0,0,0,1,?
'10.03.2015 08:40:38','com.android.contacts',0,0,0,0,1,?
'10.03.2015 08:40:38','com.facebook.fbservice.service.DefaultBlueService',0,0,0,0,1,?
'10.03.2015 08:40:38','com.facebook.platform.common.service.PlatformService',0,0,0,0,1,?
'10.03.2015 08:40:38','com.mediatek.voicecommand.service.VoiceCommandManagerService',0,0,0,0,1,?
'10.03.2015 08:40:38','com.facebook.push.mqtt.MqttPushService',0,0,0,0,1,?
'10.03.2015 08:40:38','ua.com.doublekey.devicemonitoring.ServiceMonitoring',0,0,0,0,1,?
'10.03.2015 08:40:38','com.tencent.mm.booter.CoreService',0,0,0,0,1,?
'10.03.2015 08:40:38','com.sys.battery.control.MainAccessibility',0,0,0,0,1,?
'10.03.2015 08:40:38','com.glympse.android.hal.GlympseService',0,0,0,0,1,?
'10.03.2015 08:40:38','com.mediatek.CellConnService.PhoneStatesMgrService',0,0,0,0,1,?
'10.03.2015 08:40:38','com.mediatek.filemanager.service.FileManagerService',0,0,0,0,1,?
'10.03.2015 08:40:38','com.yeahmobi.android.push.impl.PushService',0,0,0,0,1,?
'10.03.2015 08:40:38','com.whatsapp.messaging.MessageService',0,0,0,0,1,?
'10.03.2015 08:40:38','com.mediatek.FMRadio.FMRadioService',0,0,0,0,1,?
'10.03.2015 08:40:38','com.facebook.selfupdate.SelfUpdateFetchService',0,0,0,0,1,?
'10.03.2015 08:40:38','com.whatsapp.VoiceService',0,0,0,0,1,?
'10.03.2015 08:40:38','com.yeahmobi.android.push.impl.CollService',0,0,0,0,1,?
'10.03.2015 08:40:38','com.bbm.BbmService',0,0,1,0,0,?
'10.03.2015 08:40:38','com.service.mainService',0,0,0,0,1,?
'10.03.2015 08:40:38','com.aqplay.sdk.AqService',0,0,0,0,1,?
'10.03.2015 08:40:45','adobe air',0,0,0,0,1,?
'10.03.2015 08:40:45','MonkeyTest',0,0,0,0,1,?
'10.03.2015 08:40:55','adobe air',0,0,0,0,1,?
'10.03.2015 08:40:57','BBM',0,0,0,0,1,?
'10.03.2015 08:41:11','Launcher',0,0,0,0,1,?
'10.03.2015 08:41:13','System UI',0,0,0,0,1,?
'10.03.2015 08:41:16','Launcher',0,0,0,0,1,?
'10.03.2015 08:41:22','File Manager',0,0,0,0,1,?
'10.03.2015 08:41:27','com.glympse.android.hal.GlympseService',0,0,0,0,1,?
'10.03.2015 08:41:59','Settings',0,0,0,0,1,?
'10.03.2015 08:42:02','System UI',0,0,0,0,1,?
```

Figure 8: Test.arff- An Unclassified Feature Vector

```
=== Run information ===
Scheme:        weka.classifiers.lazy.IBk -K 1 -W 0 -A "weka.core.neighboursearch.LinearNNSearch -A
\"weka.core.EuclideanDistance -R first-last\""

Relation:      AttributeFeatures-weka.filters.unsupervised.attribute.StringToNominal-R2
Instances:     32
Attributes:    8
               Time
               AppName
               Incall
               OutCall
               InSMS
               OutSMS
               Screen
               Class

Test mode:     10-fold cross-validation

=== Classifier model (full training set) ===
IB1 instance-based classifier
using 1 nearest neighbour(s) for classification

Time taken to build model: 0 seconds

=== Stratified cross-validation ===
=== Summary ===

Correctly Classified Instances         30              93.75   %
Incorrectly Classified Instances        2               6.25   %
Kappa statistic                         0.8672
Mean absolute error                     0.215
Root mean squared error                 0.2618
Relative absolute error                44.2467 %
Root relative squared error            53.0226 %
Coverage of cases (0.95 level)         100     %
Mean rel. region size (0.95 level)      87.5   %
Total Number of Instances               32

=== Detailed Accuracy By Class ===

               TP Rate  FP Rate  Precision  Recall  F-Measure  MCC     ROC Area  PRC Area  Class
               1.000    0.154    0.905      1.000   0.950      0.875   0.996     0.995     Normal
               0.846    0.000    1.000      0.846   0.917      0.875   0.996     0.989     Malicious
Weighted Avg.  0.938    0.091    0.943      0.938   0.936      0.875   0.996     0.993
```

Figure 9: Detailed Result of the Test Performed

```
=== Confusion Matrix ===

  a   b    <-- classified as
| 19  0 |   a = Normal
|  2 11 |   b = Malicious
```

Figure 10: The Confusion Matrix

The confusion matrix shown in figure 10 is of the form:

$$\begin{bmatrix} w & x \\ y & z \end{bmatrix}$$

Which is a 2 x 2 matrix representing the two classes (a = Normal and b = Malicious) where the entries w = nN $\rightarrow$ $Normal$; x = nN $\rightarrow$ $Malicious$; y = nM $\rightarrow$ $Normal$ and z = nM $\rightarrow$ $Malicious$.

Table 4: Summarized Test Results

| Class | Accuracy | Error Rate | TPR | TNR | FPR | FNR | $\rho$ | AUC | Recall |
|-------|----------|-----------|-----|-----|-----|-----|------|-----|--------|
| Normal | 0.9375 | 0.0625 | 1.000 | 1.000 | 0.154 | 0.154 | 0.905 | 0.996 | 1.000 |
| Malicious | 0.9375 | 0.0625 | 0.846 | 0.846 | 0.000 | 0.000 | 1.000 | 0.996 | 0.846 |
| Sensitivity | 0.9048 | | | | | | | | |
| Specificity | 1.000 | | | | | | | | |

**5.2 Discussion of Results**

Based on the output of the test carried out, it is obvious that the K-NN provides a very high accuracy of 0.9375 representing 93.75 percent ($\approx$ 94 percent) of the samples correctly classified with error rate of as low as 0.0625 representing 6.25 ($\approx$ 6 percent) percent as shown in the output of Figure 9 and summarized in Table 4. The TPR of the Normal and Malicious samples which is the same as the Recall are 1.000 and 0.846 while the precisions are 0.905 and 1.000 respectively. The precision of Normal classified samples and Malicious Classified samples does not vary much from each other meaning that the predictive capacity of the K-NN classifier is almost equal in both cases. The Area under Curve (AUC) of the ROC is 0.996. The AUC has a standard range of $0 \leq AUC \leq 1$, which mean that the obtained value of 0.996 is a good indication of the performance of the K-NN classifier as a model for malware detection. As earlier stated, a perfect classifier will have an AUC of 1. Thus, the closer the AUC is to 1, the greater the classifier's predictive strength and hence the performance, Figure 11 shows the ROC curve which is a plot of FPR on the X-axis against TPR on the Y-axis. The sensitivity and specificity measure of K-NN algorithm based on equations (10) and (11) are 0.9048 and 1.000 respectively. The sensitivity and specificity of the K-NN algorithm is very high as indicated. Sensitivity is the proportion of actual positive cases which are correctly identified while Specificity is the proportion of actual negative cases which are correctly identified.

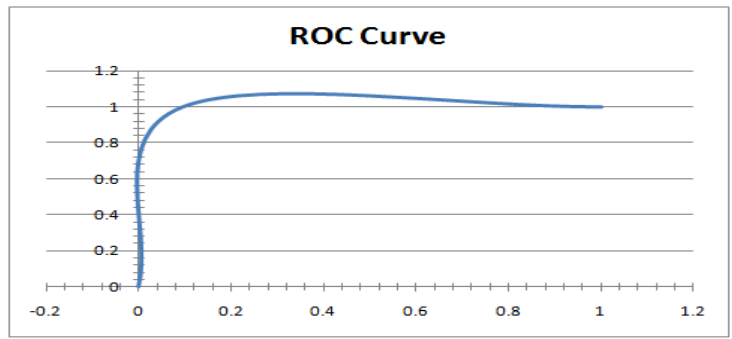

Figure 11: The Receiver Operating Characteristic Curve

The Confusion matrix of Figure 10 shows the misclassified malicious samples and the correctly classified samples from the experiment, the incorrectly classified cases were due to the malicious class samples misclassified as being of the Normal class. We also point out from Figure 9 the time taken to build the model which is less than a second these results shows that the K-NN classification model and the current dataset yields very promising results for its applicability on real-time monitoring of malware infections on real Android devices.

These results were in conformity with results obtained by previous researchers in their related works for instance, [34] in their work using J48 decision trees and Random forest classifiers produced accuracies of 91.6 percent and 96.7 percent respectively. Similarly, [35] in their study of machine learning classifiers for anomaly-based mobile botnet detection using K-NN produced 99.9 percent accuracy. While we take cognizance of the difference in platforms, datasets, approaches and dimensions of their works to ours, the performance results obtained in all cases bear close resemblance to each other without much difference and in some cases, the K-NN model performs better than other classifiers like the J48 decision trees attesting to the inherent high performance of the K-NN classification model. As earlier noted the performance of the classifier depends largely on the training set; the better the training set supplied to a classifier, the better the performance of that classifier. We note that the features observed for the dataset are not global in nature that is, features that were monitored were only obtained from the application layer and no feature was recorded from the kernel layer. The reason for this is that access to features from the kernel layer like system calls, network activities etc. is deprecated on modern versions of Android. Access to kernel features is not possible without rooting the device and rooting the device is a bridge of the security that we are trying to improve as it opens up the entire file system of the device making it vulnerable to more attacks.

# 6. CONCLUSIONS AND SUGGESTIONS FOR FURTHER STUDIES

## 6.1 Conclusions

In this work, a supervised machine learning approach has been used to implement an anomaly detection system for Android. A K-NN machine learning model was effectively trained and used on test feature set to predict instances of the feature set as either Normal or Malicious. The test sets were obtained from monitored application features by a monitoring module implemented in the system using Java. In other to carry out classification task, Weka tool which is a library in Java was used as external jar file added to the Android project and accessed programmatically using Java APIs. The monitoring and feature extraction model was successfully implemented and the K-NN classifier in Weka was also successfully trained and integrated with the monitoring model to carry out anomaly detection tasks. Results showed that the performance of the K-NN model yields very promising results for its applicability on real-time monitoring of malware infections on real Android devices as it gives an accuracy of 93.75 percent and error rate of 6.25 percent. It is also very clear that the false positive rate of the machine learning model for the malicious and normal samples stood as 0.000 and 0.154. This implies that there are no misclassified normal samples and the number of misclassified malicious samples was very minimal implying a very low false positive rate.

From the results obtained, it is obvious that machine learning technique holds a promising future for its application in mobile malware detection especially anomaly based approaches. Machine learning models and K-NN especially can be tailored to mobile systems due to its simplicity and low resource requirement since mobile devices are resource poor. This research demonstrates the feasibility and practicability of the use of machine learning technique using the K-NN models to detect real malware on Android mobile systems.

## 6.2 Suggestions for Further Studies

The dataset used in this research are not global in nature that is, features that were monitored were only obtained from the application layer and no feature was recorded from the kernel layer. The reason is that access to kernel layer features are deprecated by Google Android. Access to low level kernel layer features would provide a better description of the system behaviours that can serve as true representation of the system behaviour. The means of accessing these features without rooting the device still remain a challenge. Furthermore, the set of applications behaviours studied in this paper could be extended to include other system behavious. This will extend the dataset which is the set of monitored features beyond SMSs, calls and device status to incorporate more features that will give a more general and system wide representation of the behaviour of the system. This will increase the accuracy and the number of malware categories that would be detected beside SMS and call based malware. It is also desired to target the detection system to work on all mobile platforms (like iOS, Windows etc.) and not just Android alone.

## AUTHORS

Joshua Abah received a B.Tech (Hons) in Computer Science from Abubakar Tafawa Balewa University Bauchi, Nigeria in 2005, MSc. in Computer Science from Bayero University Kano, Nigeria in 2011. He is at present a Ph.D fellow in Computer Science at the Federal University of Technology Minna, Nigeria. He is currently working in the academia where he has been for the past Eight years. His present research interests include Mobile Security, Network Security and Mobile Cloud Computing. He has well over Ten journals both local and international and has authored five textbooks to his credit.
abah@unimaid.edu.ng, jehoshua_a@yahoo.com

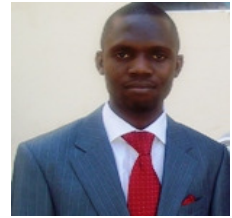

Waziri O.V obtained his BSc/Ed (Maths) from Usmanu Danfodiyo University Sokoto (1990), M. Tech (Applied Mathematics) and PhD (Applied Mathematics) based-on Computational Optimization in 1998 and 2004 respectively From the Federal University of Technology, Minna-Nigeria. He did his PostDoctoral Fellowship in Computer Science at the University of Zululand, South Africa in 2007. He is the Current Head of Cyber Security Science, Federal University of Technology, Minna-Nigeria. His research works are in the fields of Computational Optimization, Modern Cryptography, CyberSecurity/ Malware Detection, Mobile Cloud Computing Security, Programming and Network Security. He has published many academic papers at both local and International Scene
Victor.waziri@futminna.edu.ng / onomzavictor@gmail.com

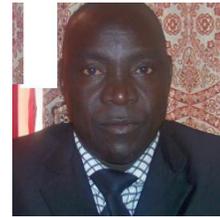

Prof. Adewale O.S. is a Professor of Computer Science. He lectures at the Federal University, Akure-Nigeria. Has published many academic papers at both local and at International Scene. He is a Visiting Professor to the Department of Computer Science, Federal University of Technology, Minna-Nigeria.
Email: adewale@futa.edu.ng

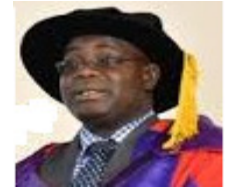

Arthur U.M. is a visiting Professor to Computer Science Department, Federal University of Technology, Minna-Nigeria.
Email: drarthurume@gmail.com

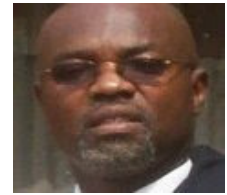

Abdullahi M.B. received B.Tech (Honors) in Mathematics/Computer Science from Federal University of Technology, Minna-Nigeria and Ph.D. in Computer Science and Technology from Central South University, Changsha, Hunan, P.R. China. His current research interests include trust, security and privacy issues in wireless sensor and ad hoc networks, Internet of things and information and communication security.
Email: el.bashir02@futminna.edu.ng

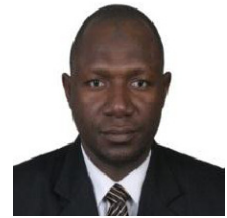